\documentclass[conference]{IEEEtran}
\IEEEoverridecommandlockouts
\usepackage{amsmath,amssymb,amsfonts}
\usepackage{graphicx}
\usepackage{textcomp}
\def\BibTeX{{\rm B\kern-.05em{\sc i\kern-.025em b}\kern-.08em
    T\kern-.1667em\lower.7ex\hbox{E}\kern-.125emX}}
    
    \usepackage{graphicx}      
\usepackage{natbib}        

\usepackage[utf8]{inputenc} 
\usepackage[english]{babel} 

\graphicspath{ {./img/} }

\usepackage[font=small]{caption}

\usepackage{chngcntr}

\usepackage[table,xcdraw]{xcolor}

\usepackage{booktabs} 
\usepackage{url} 
\usepackage[bookmarks,unicode,hidelinks]{hyperref}

\usepackage{enumitem}
\usepackage[ruled]{algorithm}
\usepackage{algpseudocode}
\usepackage{multicol}

\newtheorem{theorem}{Theorem}[section]
\newtheorem{corollary}{Corollary}[theorem]

\newtheorem{definition}{Definition}[section]
\newtheorem{remark}{Remark}

\newtheorem{pitfall}{Pitfall}[theorem]
\newtheorem{fallacy}{Fallacy}[theorem]

\begin{document}

\title{A new metric for evaluating the performance and complexity of computer programs \\
{\footnotesize A new approach to the traditional ways of measuring the complexity of algorithms and estimating running times}
}

\author{\IEEEauthorblockN{Rares FOLEA}
\IEEEauthorblockA{\textit{Computer Science \& Engineering Department} \\
\textit{Faculty Of Automatic Control And Computers} \\
\textit{University Politehnica Of Bucharest}\\
Bucharest, Romania \\
rares.folea@stud.acs.upb.ro}

\and

\IEEEauthorblockN{Emil-Ioan SLUSANSCHI}
\IEEEauthorblockA{\textit{Computer Science \& Engineering Department} \\
\textit{Faculty Of Automatic Control And Computers} \\
\textit{University Politehnica Of Bucharest}\\
Bucharest, Romania \\
emil.slusanschi@cs.pub.ro}
}

\maketitle

\begin{abstract}
This paper presents a refined complexity calculus model: r-Complexity, a new asymptotic notation that offers better complexity feedback for similar programs than the traditional Bachmann-Landau notation, providing subtle insights even for algorithms that are part of the same conventional complexity class. The architecture-dependent metric represents an enhancement that provides better sensitivity with respect to discrete analysis.
\end{abstract}


\section{Introduction}

This paper will present a new approach in the field of algorithm's computational complexity. This new calculus model aims to produce new asymptotic notations that offer better complexity feedback for similar algorithms, providing subtle insights even for algorithms that are part of the same conventional complexity class $\Theta(g(n))$ denoted by an arbitrary function $g:\mathbb{N}\longrightarrow\mathbb{R}$, in the definition of \textit{Bachmann–Landau} (B-L) notations~\cite{giumale2004introducere}. The additional information contained by \textit{r-Complexity} classes consists of the fine-granularity obtained by the model based on a refined clustering strategy for functions that used to belong to the same \textit{B-L} group in different complexity classes, established on asymptotic constant analysis.

\textbf{Section \ref{sect:model}} defines the \textit{r-Complexity} model, an asymptotic notation, expressed as function $f:\mathbb{N}\longrightarrow\mathbb{R}$. The function is characterized by the size of the input, while the evaluated value $f(n)$, for a given input size $n$, represents the amount of resources needed in order to compute the desired result.

\textbf{Section \ref{sect:algs}} outlines the connection between this model and various computer algorithms, offering a set of metrics for comparing algorithms interval-based and asymptotic performances. 

\textbf{Section \ref{sect:practice}} presents automatic methods for estimating the associated r-Complexity for a generic algorithm, while \textbf{Sections \ref{sect:matrix}} and \textbf{\ref{sect:chess}} present two use-cases for applying this model: matrix multiplication and optimization techniques as well as an analysis on perfect chess algorithms.

\section{A refined complexity calculus model}
\label{sect:model}

The following notations and names \cite{mogocs2015new} will be used for describing the asymptotic behavior of a algorithm's complexity characterized by a function, $f:\mathbb{N}\longrightarrow\mathbb{R}$. \\
We define the set of all complexity calculus $\mathcal{F}= \lbrace f:\mathbb{N}\longrightarrow\mathbb{R} \rbrace$
\\Assume that $n, n_{0}\in\mathbb{N}$. Also, we will consider an arbitrary complexity function $g \in \mathcal{F}$.
Acknowledge the following notations $\forall r \neq 0$:

\begin{definition}
    \textbf{Big \textit{r-}Theta}: This set defines the group of mathematical functions similar in magnitude with  $g(n)$ in the study of asymptotic behavior. A set-based description of this group can be expressed as:
    \[\begin{split}
          \Theta_{r}(g(n)) = \lbrace f \in \mathcal{F}\ |\ \forall c_{1}, c_{2} \in \mathbb{R}^{*}_{+} \ s.t. c_{1}< r < c_{2} , \exists n_{0} \in \mathbb{N}^{*}\ \\ s.t.\ \ c_{1} \cdot g(n) \leq f(n) \leq c_{2} \cdot g(n)\ ,\  \forall n \geq n_{0} \rbrace
    \end{split} \]
\end{definition}

\begin{definition}
    \textbf{Big \textit{r-}O}: This set defines the group of mathematical functions that are known to have a similar or lower
    asymptotic performance in comparison with  $g(n)$. The set of such functions is defined as it follows:
    \[\begin{split} \mathcal{O}_{r}(g(n)) = \lbrace f \in \mathcal{F}\ |\ \forall c  \in \mathbb{R}^{*}_{+} \ s.t.\  r<c, \exists n_{0} \in \mathbb{N}^{*}\ \\ s.t.\  f(n) \leq c \cdot g(n),\  \forall n \geq n_{0} \rbrace \end{split}\]
\end{definition}

\begin{definition}
    \textbf{Big \textit{r-}Omega}: This set defines the group of mathematical functions that are known to have a similar or higher asymptotic performance in comparison with  $g(n)$. The set of all function is defined as:
    \[ \begin{split} \Omega_{r}(g(n)) = \lbrace f \in \mathcal{F}\ |\ \forall c  \in \mathbb{R}^{*}_{+}\ s.t. \ c < r, \exists n_{0} \in \mathbb{N}^{*}\ \\ s.t.\  f(n) \geq c \cdot g(n),\  \forall n \geq n_{0} \rbrace \end{split}\]
\end{definition}

\begin{definition}
    \textbf{Small \textit{r-}O}:
    This set defines the group of mathematical functions that are known to have a humble
    asymptotic performance in comparison with  $g(n)$. The set of such functions is defined as it follows:
    \[\begin{split} o_{r}(g(n)) = \lbrace f \in \mathcal{F}\ |\ \forall c \in \mathbb{R}^{*}_{+}, \exists n_{0} \in \mathbb{N}^{*}\ \\ s.t.\  f(n) < c \cdot g(n),\  \forall n \geq n_{0} \rbrace \end{split}\]
\end{definition}
\begin{remark}
    This set is defined for symmetry of the model and it is equal with the set defined by \textbf{Small O} notation in \textit{B-L notations}, as the definition is independent on $r$.
\end{remark}

\begin{definition}
    \textbf{Small \textit{r-}Omega}:
    This set defines the group of mathematical functions that are known to have a commanding asymptotic performance in comparison with  $g(n)$.
    The set of such functions is defined as it follows:
    \[\begin{split} \omega_{r}(g(n)) = \lbrace f \in \mathcal{F}\ |\ \forall c \in \mathbb{R}^{*}_{+}, \exists n_{0} \in \mathbb{N}^{*}\ \\ s.t.\  f(n) > c \cdot g(n),\  \forall n \geq n_{0} \rbrace \end{split}\]
\end{definition}
\begin{remark}
    This set is defined for symmetry of the model and it is equal with the set defined by \textbf{Small Omega} notation in \textit{B-L notations}, as the definition is independent on $r$.
\end{remark}


An interesting property of the \textbf{Big \textit{r-}Theta},  \textbf{Big \textit{r-}O} and \textbf{Big \textit{r-}Omega} classes is the simple technique of conversion between various values for the $r$s parameters. The following results arise:

\begin{theorem}
    \textbf{Big r-Theta conversion}:
    \[  f \in \Theta_{r}(g) \Rightarrow f \in \Theta_{q} \left( \frac{q}{r} \cdot g \right) \ \forall r,q \in \mathbb{R}_{+}\]
\end{theorem}

Another interesting result is obtained by multiplying the last equation by $\frac{r}{q}$: \\
$\forall c_{1}, c_{2} \in \mathbb{R}^{*}_{+} \ s.t. c_{1} < q < c_{2} , \exists n_{0}' = n_{0} \in \mathbb{N}^{*}\ \\ s.t.\ \frac{r}{q} \cdot \frac{q}{r}\cdot c_{2} \cdot g(n)\ \leq \frac{r}{q} \cdot f(n) \leq \frac{r}{q} \cdot \frac{q}{r}\cdot c_{2} \cdot g(n)\ ,\  \forall n \geq n_{0}$
\begin{corollary}
    \[ \frac{r}{q} \cdot f \in \Theta_{q} \left( g \right)\].
\end{corollary}

\begin{theorem}
    \textbf{Big r-O Conversion}:
    \[  f \in \mathcal{O}_{r}(g) \Rightarrow f \in \mathcal{O}_{q} \left( \frac{q}{r} \cdot g \right) \ \forall r,q \in \mathbb{R}_{+}\]
\end{theorem}
\begin{corollary}
    The following conversion relationship arises:
    \[  f \in \mathcal{O}_{r}(g) \Rightarrow \frac{q}{r} \cdot f \in \mathcal{O}_{q} \left(  g \right) \ \forall r,q \in \mathbb{R}_{+}\]
\end{corollary}
\begin{theorem}
    \textbf{Big r-Omega Conversion}:
    \[  f \in \Omega_{r}(g) \Rightarrow f \in \Omega_{q} \left( \frac{q}{r} \cdot g \right) \ \forall r,q \in \mathbb{R}_{+}\]
\end{theorem}
\begin{corollary}
    The following conversion relationship arises:
    \[  f \in \Omega_{r}(g) \Rightarrow \frac{q}{r} \cdot f \in \Omega_{q} \left( g \right) \ \forall r,q \in \mathbb{R}_{+}\]
\end{corollary}

For Big notations $(\Theta, \mathcal{O}, \Omega)$, we present further some results:

\begin{theorem}
    Relationship between Big r-Theta and Big Theta:

Consider $f$ a continuous functions and let $\exists r \in \mathbb{R}_{+}$.

    \[  f \in \Theta_{r}(g) \Rightarrow f \in \Theta (g) \]
    \[  f \in \Theta(g) \Rightarrow \exists r \in \mathbb{R}_{+}\ f \in \Theta_{r}(g) \]
\end{theorem}

\begin{theorem}
    Relationship between Big r-O and Big O:
    \[  f \in \mathcal{O}_{r}(g) \Rightarrow f \in \mathcal{O} (g) \]
    \[  f \in \mathcal{O}(g) \Rightarrow \exists r \in \mathbb{R}_{+}\ f \in \mathcal{O}{r}(g) \]
\end{theorem}

\begin{theorem}
    Relationship between Big r-Omega and Big Omega:
    \[  f \in \Omega_{r}(g) \Rightarrow f \in \Omega (g) \]
    \[  f \in \Omega(g) \Rightarrow \exists r \in \mathbb{R}_{+}\ f \in \Omega_{r}(g) \]
\end{theorem}

\section{Algorithms and Complexity}
\label{sect:algs}

\subsection{Estimating computational time based on Normalized r-Complexity}
Let an arbitrary algorithm $Alg$ characterized by the complexity function $f$ with a variable input dimension $n \in \mathcal{N}^{*}$. Consider that the input size is bounded such that $n \in [n_{min}, n_{max}]$. \\
We aim to define various metrics for approximation an average computational time required based on the size of the input and the algorithm's complexity function  $T(n_{min}, n_{max})$.\\

\begin{definition}
    \textbf{RM1}

    Defined as a metric for time estimation (capable of generalization to any other estimators) based on arithmetic mean in Normalized r-Complexity model is defined as follows:
    \[  T(n_{min}, n_{max}) = \dfrac{\sum\limits_{n=n_{min}}^{n_{max}} g_{1}(n)}{n_{max} - n_{min} + 1}  \]
\end{definition}

\begin{definition}
    \textbf{RM2}

    Defined as a metric for time estimation (capable of generalization to any other estimators) based on Mean-Value Theorem (Lagrange) using integrals in Normalized r-Complexity model is defined as follows:
    \[  T(n_{min}, n_{max}) = \dfrac{\int\limits_{n_{min}}^{n_{max}} g_{1}(n) dn}{n_{max} - n_{min}}  \]
\end{definition}

The previous two metrics are tailored for systems where the input size is bounded but there is no additional knowledge regarding the weights and probabilities of occurrence. If this information is available, we can redefine the previous metrics using the acquisition data.

\begin{definition}
    \textbf{ERM1}, an enhanced metric for time estimation based on arithmetic mean in Normalized r-Complexity model is defined as follows:
    \[  T(n_{min}, n_{max}) = \sum\limits_{n=0}^{f} p_{n} \cdot g_{1}(n + n_{min})  \]
    where:
    \begin{itemize}
        \item $p_{0}$ is the weight associated with $n_{0} = n_{min}$
        \item $p_{1}$ is the weight associated with $n_{1} = n_{min + 1}$
        \item $p_{f}$ is the weight associated with $n_{f} = n_{max}$
    \end{itemize}
    and $f = max - min + 1$.
\end{definition}

\begin{definition}
    \textbf{ERM2}, an enhanced metric for time estimation based on Mean-Value Theorem (Lagrange) using integrals in Normalized r-Complexity model is defined as follows:
    \[  T(n_{min}, n_{max}) =\sum\limits_{k=0}^{f-1} p_{k} \cdot \int\limits_{n_{k}}^{n_{k+1}} g_{1}(n) dn  \]
    where:
    \begin{itemize}
        \item $p_{0}$ is the weight associated with the \textit{probability} of the input to be bounded in the interval $[n_{0}, n_{1}]$
        \item $p_{1}$ is the weight associated with the \textit{probability} of the input to be bounded in the interval $[n_{1}, n_{2}]$
        \item $p_{f-1}$ is the weight associated with the \textit{probability} of the input to be bounded in the interval $[n_{f-1}, n_{f}]$
    \end{itemize}
    and $f = max - min + 1$, and $n_{0} = n_{min}, n_{f} = n_{max}$.
\end{definition}

\subsection{Comparing algorithms interval-based performances}
Consider an application responsible for scheduling a football league agenda for the next competitive season, avoiding conflicts and following specific objectives. This problem can be modeled and solved as a \textit{constraint satisfaction problem} (CSP) with different flavors. An asymptotic performance analyzer would simply pick the lowest complexity function in consideration to asymptotic behavior. However, the application is not designed to run on extremely large input size, as the cardinal of the set of all teams part of a football league is a bounded well-known small integer (\textit{most leagues have between 14 and 20 teams}). Therefore, it may be a wise choice to have another method of comparing different algorithms with respect to finite upper bounded input. Asymptotic performance is thus not always relevant for computer programs that have a settled range of input sizes in order to solve a specific task, or an interval-based approximation, with or without weights on sub-specific intervals.

To address this issue we propose the subsequent Theorem, Corollary and Remarks:

\begin{theorem}
    If $ \dfrac{T_{1}(n_{min}, n_{max})}{T_{2}(n_{min}, n_{max})} = r \in [0,1) $ , then $Alg1$ will terminate faster (in average) than $Alg2$ for a probabilistic distribution of input size $n \in [n_{min}, n_{max}]$, assuming ERM2 definitions for $ T_{1}(n_{min}, n_{max})$.
\end{theorem}

\begin{corollary}
    If $ \dfrac{T_{1}(n_{min}, n_{max})}{T_{2}(n_{min}, n_{max})} = r \in (1,\infty) $ , then $Alg2$ will terminate faster (in average) than $Alg1$ for a probabilistic distribution of input size $n \in [n_{min}, n_{max}]$.
\end{corollary}

\begin{remark}
    If  $ \dfrac{T_{1}(n_{min}, n_{max})}{T_{2}(n_{min}, n_{max})} = 1$, r-Complexity model considers $Alg1$ and $Alg2$ equivalent from a computational cost-based perspective for probabilistic distribution of input size $n \in [n_{min}, n_{max}]$.
\end{remark}

\begin{remark}
    Theorem stands likewise using any metric defined for $T_{1}(n_{min}, n_{max})$. (\textbf{RM1}, \textbf{RM2}, \textbf{ERM1}).
\end{remark}

\section{r-Complexity}
\label{sect:practice}

\subsection{Human-driven calculus of r-Complexity}
The associated r-Complexity class can be calculated, by hand, for any given algorithm, provided a predefined instruction set architecture and the correspondence between generic instructions and required time for execution as well as enhanced hardware designs details related to the total execution time (number of stages of pipeline, scalability degree, etc.). Even if the process of calculating an exact r-Complexity class associated to a real algorithm is unpractical, the method provided can be applied with colossal endeavor.

\begin{remark}
    An example, calculated for a naive algorithm solving exhaustively the N-Queens’ Problem would have the r-Complexity function associated with the algorithm $(408 \cdot n^{2} \cdot n!)$. This is from the same tradition complexity class $ O(n^2\cdot n!)$.
\end{remark}

\subsection{Automatic estimation of r-Complexity}
This section aims to present a solution for automation for calculating an approximate of the associated r-Complexity class for any given algorithm. The prerequisites for this method implies a technique for obtaining relevant metric-specific details for diversified input dimensions. For instance, if time is the monitored metric, there must exist a collection of pertinent data linking the correspondence between input size and the total execution time for the designated input size.

\subsection{Estimation for algorithms with known B-L Complexity}

\begin{figure*}
\begin{multicols}{3}
    \includegraphics[width=\linewidth]{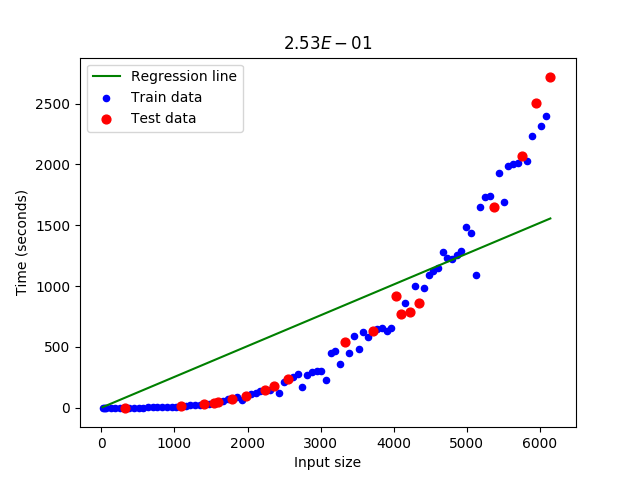}\par 
    \includegraphics[width=\linewidth]{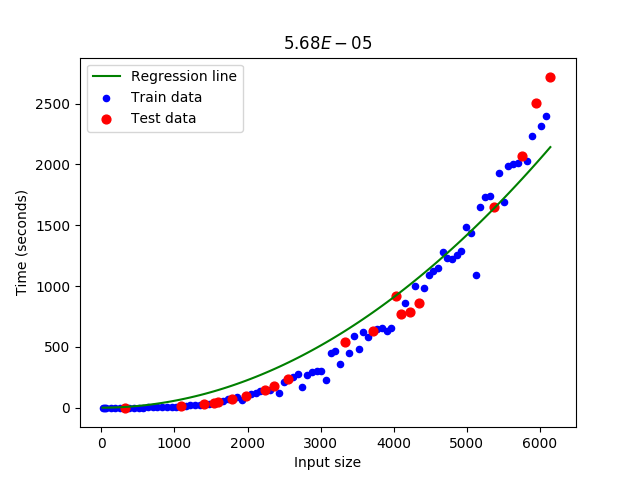}\par 
    \includegraphics[width=\linewidth]{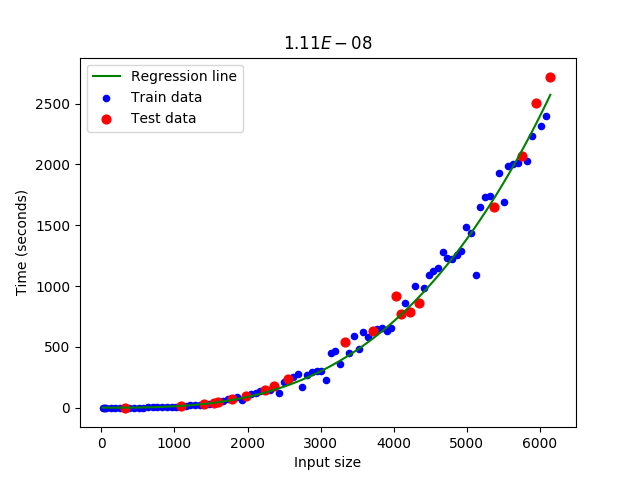}\par
    
\end{multicols}
\caption{Various prediction boundaries based on accommodated training dataset using multiple relations $g$ $\Rightarrow$ ($n^1, n^2, n^3$). Training data are obtained for different input size for a naive matrix multiplication algorithm in $\mathbb{O}(n^{3})$}
\label{fig:ml-trick}
\end{figure*}

Reckoning an associated r-Complexity class ($f$) for an algorithm with established B-L Complexity ($g$) consists in the process of tailoring an suitable constant $c$, such that $f \approx \Theta_{1}(c \cdot g)$ or in Big-O calculus, $f \leq  \mathbb{O}_{1} (c \cdot g)$. The approach presented below is a particularized version of linear regression, which attempts to model the relationship between various variables by fitting a linear equation to observed data. Even if the model generally follows the classical pattern of a Machine Learning Process (training, predicting, etc.), where a training example consists of a pair $(inputSize, metricValue)$.

A trick (frequently used in data science) is used to adjust the entry values if the B-L relationship between the $inputSize$ and the metric is known. In order to adjust the learning set to a more knowledgeable set, we can extract new features and transform all the ($inputSize$, $metricValue$) pairs into ($g(inputSize)$, $metricValue$), where $g$ is the known B-L Complexity function converted into Normal form.

The importance of this trick can be emphasized comparing the classical linear regression model with various learning datasets. For the matrix multiplication problem, a naive algorithm (with B-L Complexity $\mathbb{O}(n^{3})$) has been implemented. After testing, the algorithm has been deployed and executed matrix multiplications for various sizes of the matrices. As an intuition (due to the associated complexity function $O(n^3)$ in the B-L Complexity model), the natural fit, as seen in Figure~\ref{fig:ml-trick} was obtained when using $g(n) = n^3$ with consideration to generalization. If we choose much much bigger degree polynomial transformations, we may obtain better results on this data-sets, but the models are becoming subject to over-fit.

\subsection{Estimation for algorithms with unknown B-L Complexity}
Estimation for algorithms with unknown B-L Complexity becomes a lot more difficult as there are numerous possible candidates for a matching complexity function.

An enhanced model, based on a general polynomial performance model normal form~\cite{calotoiu2018automatic} for complexity functions, should contain an exponential behavior, which is often seen as a synergy between NP-Hard problems. Thus, we propose the following general expression:

\[ f(n) =\sum\limits_{t=1}^{y}  \sum\limits_{k=1}^{x} c_{k} \cdot n^{p_{k}} \cdot log_{l_{k}}^{j_{k}}(n) \cdot e_{t}^{n} \cdot  \Gamma(n)^{g_{k}} \]
This representation is, of course, not exhaustive, but it works in most practical schemes. An intuitive motivation is a consequence of how most computer algorithms are designed~\cite{calotoiu2018automatic}.

\section{Matrix Multiplication Use-case}
\label{sect:matrix}

\subsection{Naive and optimized implementations}
We analyzed various naive matrix multiplication algorithms $O(n^3)$ with memory-access improvements (cache-locality of loops, Blocked Matrix Multiplication) and an efficient implementation of Strassen's algorithm. Using the method described in estimating section, we can tailor an architecture-specific complexity function $ f(n)  = c \cdot n^3$. After training the regression models for each algorithm, we obtained the coefficients $c$, that defines the same complexity function.

\begin{figure}
    \centering
    \includegraphics[width=0.45\textwidth]{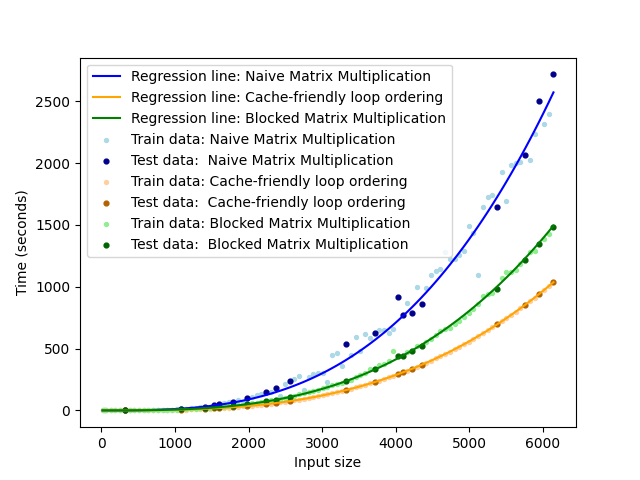}
    \caption{Regression lines corresponding to each of the matrix-multiplication algorithms.}
    \label{fig:reg-mm-algo}
\end{figure}

The results presented in Figure~\ref{fig:reg-mm-algo} have been obtained on an Intel i5 3.2GHz, x86\_64 Architecture with L1d cache: 32K, L1i cache: 32K, L2 cache: 256K, L3 cache: 6144K. We do not postulate that the methods above cannot be enhanced or that the efficiency of the optimizations are in a specific order. We aim to provide various estimation for these implementations of the matrix multiplication algorithms with known $O(n^3)$ Complexity. Please remark the natural distribution of the two cache-friendly algorithms presented on larger data-sets vs. the naive algorithm, susceptible to outliers.
    
In this representation, the complexity function is scaled to produce output in seconds. In order to obtain r-Complexity function, as shown in Table~\ref{tab:complexity-function}, multiplying with processor frequency is mandatory $HZ \approx 3.2 \cdot 10^9 $.
\begin{table}[H]
    \centering
    \begin{tabular}{|
    >{\columncolor[HTML]{FFCCC9}}l |
    >{\columncolor[HTML]{32CB00}}l |}
        \hline
        \cellcolor[HTML]{C0C0C0}\textit{\textbf{Algorithm}} & \cellcolor[HTML]{C0C0C0}{\color[HTML]{000000} \textit{\textbf{\begin{tabular}[c]{@{}l@{}}
                                                                                                                                Complexity Function
        \end{tabular}}}} \\ \hline
        \textbf{Naive Matrix Multiplication}                & {$ O_{1}(\textbf{1.109} \cdot \textbf{10}^{\textbf{-8}} \cdot \textbf{HZ} \cdot n^3) $}                                                                                      \\ \hline
        \textbf{Cache-friendly loop ordering}               & {$ O_{1}(\textbf{4.472} \cdot \textbf{10}^{\textbf{-9}} \cdot \textbf{HZ} \cdot n^3) $ }                                                                                      \\ \hline
        \textbf{Blocked Matrix Multiplication}              & {$ O_{1}(\textbf{6.441} \cdot \textbf{10}^{\textbf{-9}} \cdot \textbf{HZ} \cdot n^3) $ }                                                                                      \\ \hline
    \end{tabular}
    \caption{\label{tab:complexity-function}Architecture specific coefficients}
\end{table}
\vspace{-1em}

\subsection{Strassen's implementation}

\begin{figure}
\centering
\includegraphics[width=0.45\textwidth]{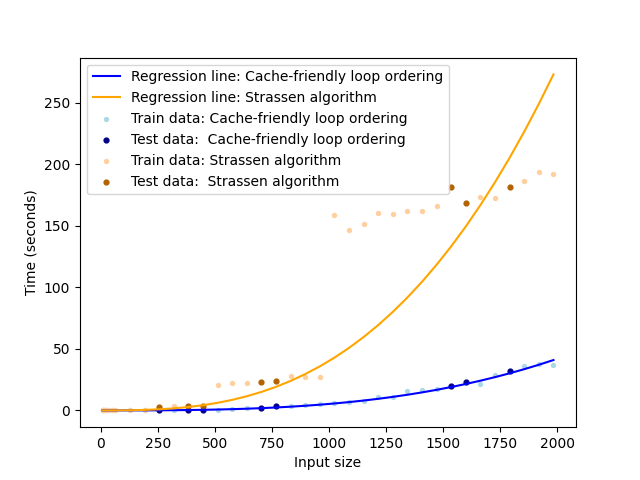}
\caption{Regression lines corresponding to Cache-friendly loop ordering matrix multiplication and Strassen's algorithm.}
\label{fig:reg-cache-strass}
\end{figure}

For a while, we will leave the $O(n^3)$ matrix multiplication algorithms and focus on a new approach. As mentioned before, Strassen's proposed a matrix multiplication algorithm with complexity $O(n^{log_{2}(7)}) \approx O(n^{2.80735})$. We aim at comparing this algorithm with the Cache-friendly loop ordering solution presented in the previous section.

In the traditional approach, without r-Complexity analysis, we could not distinguish cases in which Strassen's Algorithm could perform worse than any optimized $O(n^3)$ matrix multiplication solution.

The regression line corresponding to the Cache-friendly loop ordering matrix multiplication algorithm is shown in Figure~\ref{fig:reg-cache-strass} by $f(n) = 5.23 \cdot 10^{-9} \cdot HZ \cdot n^{3} $, while the regression line corresponding to the the Strassen's algorithm is described by $g(n) = 1.59 \cdot 10^{-7} \cdot HZ \cdot n^{2.80}$. Even if the asymptotic behavior for the Strassen's algorithm is desired in comparison with the cubic performance, for finite input size the Cache-friendly loop ordering matrix multiplication algorithm can perform better, despite $\lim_{n\to\infty} \dfrac{g(n)}{f(n)} = 0$. The nature of the non-polynomial local behaviour of the Strassen algorithm is based on architecture considerations, such as the overhead introduced by specific function calls, stack manipulations and memory allocation and management computational cost. The presented results are recorded on a x86\_64 Intel(R) Xeon(R) Gold 5218 CPU @ 2.30GHz with L1d cache: 32K, L1i cache: 32K, L2 cache: 1024K, L3 cache: 22528K, CPU max frequency: 3.9GHz.

\renewcommand\thetheorem{5}
\begin{fallacy}
Let $Alg1$ an algorithm with the complexity function $f_{1} \in \Theta(g_1(n))$  and $Alg2$ an algorithm with the complexity function $f_{2} \in \Theta(g_2(n))$. $Alg2$ must perform better than $Alg1$ \textit{for any size of input} in regard with the specified metric if $\lim_{n\to\infty} \dfrac{g_2(n)}{g_1(n)} = 0$.
\end{fallacy}

\begin{fallacy}
Adapted version for matrix multiplication:

Let $Alg1$ an algorithm with the complexity function $f_{1} \in \Theta(n^3)$  and $Alg2$ an algorithm with the complexity function $f_{2} \in \Theta(n^{2.80735})$. $Alg2$ must perform better than $Alg1$ \textit{for any size of input} in regard with the specified metric.
\end{fallacy}

Working with traditional complexity does not imply an universal increase in performance for $Alg2$, but an asymptotic comparison, while the fallacies presented in the previous statements assume an universal behavior. The correct manifest would be: $Alg2$ must perform better than $Alg1$ \textit{for sufficient large size of input} in regard with the specified metric if $\lim_{n\to\infty} \dfrac{g_2(n)}{g_1(n)} = 0$.

The collocation \textit{"Sufficient large size of input"} is essential and it means that starting from a range of input size $n_{0}$ finite, better performances are obtained using $Alg2$.

\begin{figure}
\centering
\includegraphics[width=0.45\textwidth]{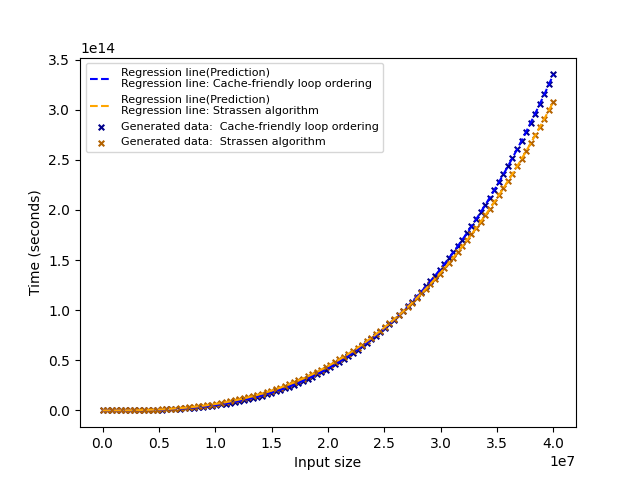}
\caption{Predictions for Cache-friendly loop ordering matrix multiplication algorithm and Strassen's algorithm.}
\label{fig:catch-point}
\end{figure}

\begin{figure*}[ht]
\begin{multicols}{2}
    \centering
    \includegraphics[width=0.9 \linewidth]{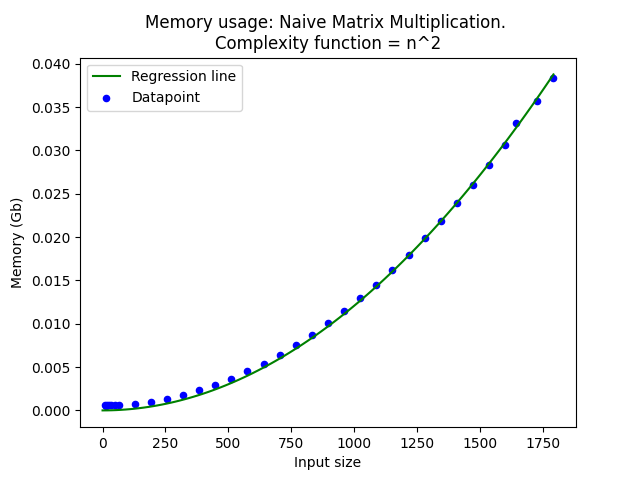}\par 
    \centering
    \includegraphics[width=0.9 \linewidth]{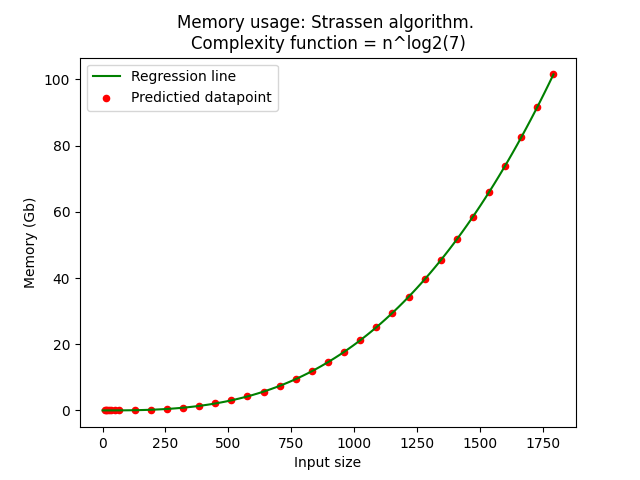}\par 
\end{multicols}
\caption{Allocated and peak memory usage for the considered algorithms.}
\label{fig:memory}
\end{figure*}

\renewcommand\thetheorem{5}
\begin{pitfall}
Let $Alg1$ an algorithm with the complexity function $f_{1} \in \Theta(g_1(n))$  and $Alg2$ an algorithm with the complexity function $f_{2} \in \Theta(g_2(n))$ and $\lim_{n\to\infty} \dfrac{g_2(n)}{g_1(n)} = 0$.

Even if the $Alg1$ may perform better than $Alg2$ for some cases, the nature of this behavior is superficial and, in general, for regular routines, $Alg2$ will still perform better.
\end{pitfall}

\begin{pitfall}
Adapted version for matrix multiplication:

Even if the Cache-friendly loop ordering matrix multiplication algorithm may perform better than the Strassen's algorithm for some cases, the nature of this behavior is superficial and, in general, for regular routines, the Strassen's algorithm will still perform better.
\end{pitfall}

The key of the previous pitfalls is the meaning of \textit{"in general, for regular routines"}, because this phrase is extremely context-dependent. The separation point where the performances of Strassen's algorithm catch up with (and overtake) the Cache-friendly loop ordering algorithm is provided by equalizing the two complexity functions:

$f(n) = 5.23 \cdot 10^{-9} \cdot HZ \cdot n^{3} $ and $g(n) = 1.59 \cdot 10^{-7} \cdot HZ \cdot n^{2.80} $ The nontrivial solution $n_{0}$ is obtained by solving $ 159 * n_{0}^{2.8} = 5.23* n_{0}^3 $, where $n_{0} \neq 0$. The solution is $n_{0} \approx 2.5 \cdot 10^7$.

For any matrix multiplication task with input size greater than $\approx 2.5 \cdot 10^7$ (25 million element matrices), better results will be obtained using Strassen's algorithm.

In Figure~\ref{fig:catch-point} one can observe that the total execution time for $25$ million element matrices, can be estimated at $g(2.5 \cdot 10^7) \approx f(2.5 \cdot 10^7) = 5.23 \cdot 10^{-9} \cdot (2.5 \cdot 10^7)^{3} \approx 8.13 \cdot 10^{13}$ seconds. This number $8.13 \cdot 10^{13} $ of seconds is the equivalent of around $25,762$ centuries.

If by "\textit{regular routines}" was meant multiplying \textit{25 million element matrices} and having the resources to await for $25,762$ centuries for the result, than Strassen's algorithm is the perfect solution for your problem. Otherwise, one should refer to a traditional approach.

\subsection{Memory considerations}

Up to this point, the only metric described in the prediction process of tailoring a complexity function was the \textbf{time} complexity. However, this is not the only resource that is important when designing algorithms and computer programs. A close match is represented by the total memory usage or peak memory usage of a computer program during the execution. Even if nowadays, memory is generally large enough to accommodate most of the possible algorithms, there are special situations in which memory management is critical.

The total peak allocated memory usage during the execution of the Cache-friendly loop ordering matrix multiplication algorithm and the total (estimated) peak memory usage for the execution of the Strassen's algorithm is shown in Figure~\ref{fig:memory}.

The problem with the Strassen's memory algorithm is that memory usage does not have a smooth improvement in growth. It varies in steps based on the powers of 2 (the cause is due to the recursion nature of the algorithm and at each iteration dividing in half). Every double in input size produces a $7x$ increase in peak memory usage. Tailoring an complexity function of type $c \cdot n^{log_2(7)}$ provides a good evaluation.

In theory, a smaller time-complexity always produce better asymptotically performances. The problems arise when we address other architectural aspects. Consider that an various algorithms uses memory usage differentiated. In order to perform precisely, a required condition is that the peak memory usage during the execution of the algorithm is at most equal with the total storage capacity of the physical RAM (ignore additional issues such as operating system memory overhead or translation concerns as well as pagination).

Consider the peak memory usage for the last two algorithms analyzed. The behavior can be tailored by the individual associated memory complexity function: \\ $f_{cache\ friendly} = 1.20 \cdot 10^{-8} \cdot n^2 $ and  $f_{strassen} = 7.89 \cdot 10^{-8} \cdot n^{2.7} $:

Recall that the critical point in order to make Strassen's algorithm perform better was estimated at around 25 million element matrices. For this value, the peak memory usage can be estimated at \textbf{7.43 ZB}. ($7.43 \cdot 10^{12}$ GB)
This amount of memory storage can be used to store \textbf{5,589,898} centuries of \textbf{1080p} digital video content \textit{(at a rate of 1.5GB/hour).} in system memory.

Using sophisticated group theory, further asymptotic improvements are provided by the Coppersmith–Winograd algorithm but these are only of theoretical interest, since the the associated r-Complexity function make these algorithms impractical~\cite{le2012faster}.

\begin{figure*}
\begin{multicols}{2}
    \includegraphics[width=\linewidth]{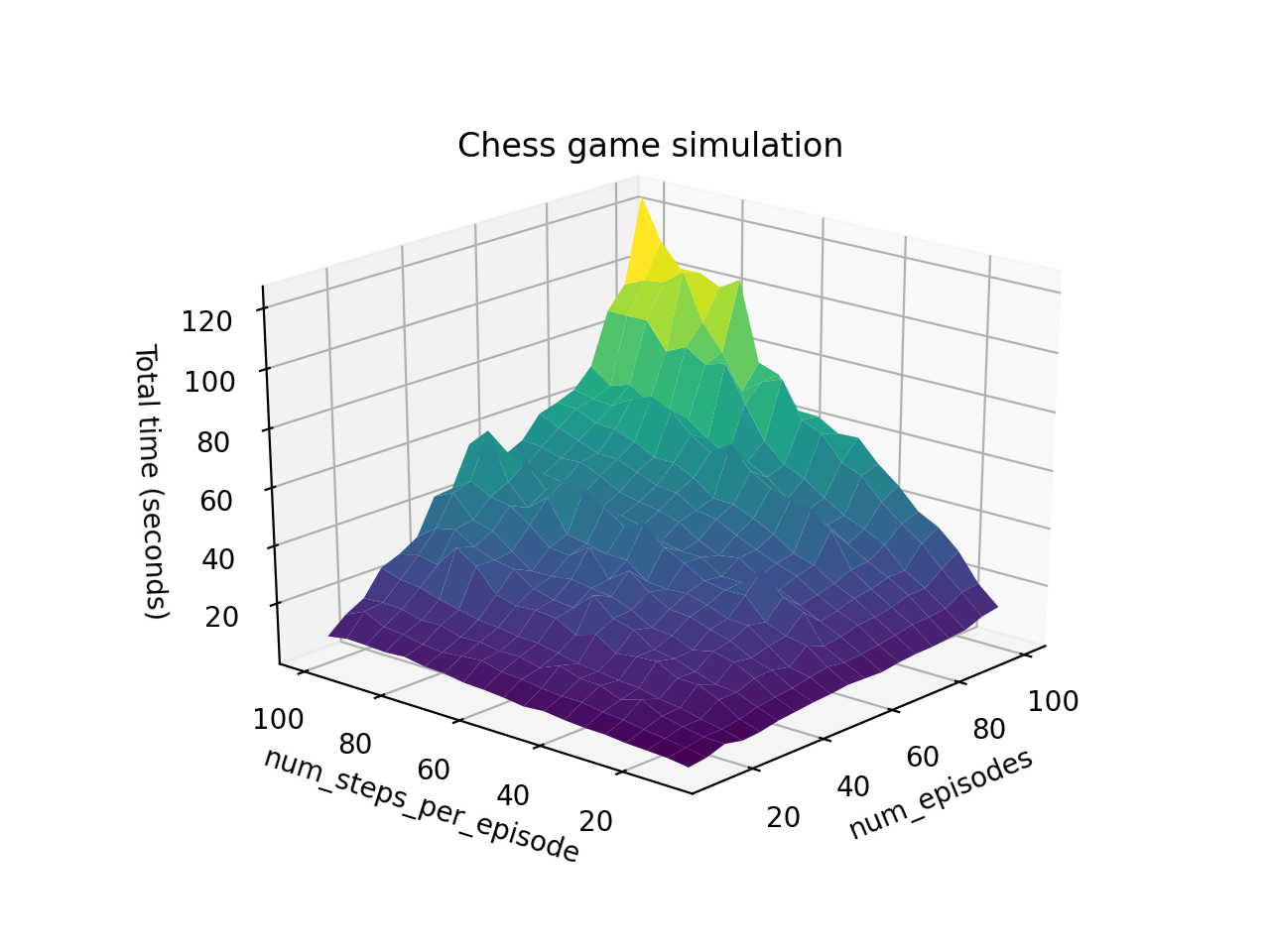}\par 
    \includegraphics[width=\linewidth]{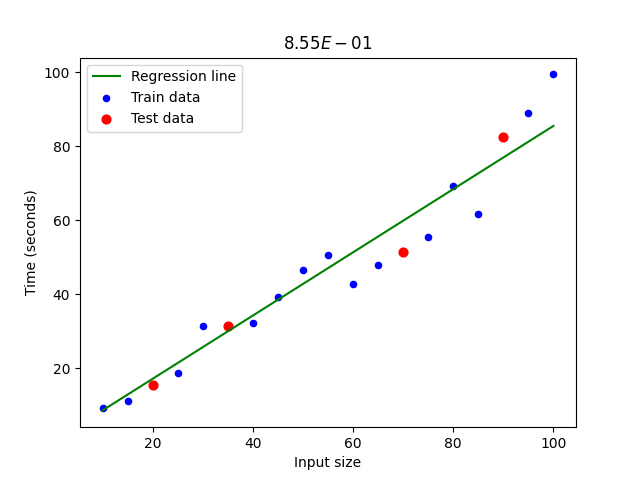}\par 
\end{multicols}
\caption{Total computing times for simulating various games of chess.}
\label{fig:computing-chess}
\end{figure*}

\begin{pitfall}
Never use Strassen's algorithm.
\end{pitfall}
All the analyzed data was obtained by analyzing a \textbf{specific} implementation of the Strassen's Matrix Multiplication Algorithm. Not all algorithmic implementation performs the same. There may be optimizing techniques to overpass some issues, especially the deep recursion problems that is raised. However, there are memory management tricks that substantially decrease the peak memory usage. In fact, this algorithm is not \textbf{galactic} and is used in practice. A galactic algorithm has the property that it is faster than other algorithm for inputs that are sufficiently large, but where sufficiently large is enormous such that that the algorithm is never used in practice~\cite{le2012faster}.

All things considered, the r-Complexity metric provides powerful insights in comparison with different complexity algorithms as illustrated by the matrix-multiplication examples.

\section{Chess-game Use-case}

\label{sect:chess}

\subsection{Computing chess}
Figure~\ref{fig:computing-chess} shows the total computing time based on the variation of parameters corresponding to the total number of games and the average length of a game. The results are obtained on a 2.3 GHz Intel Core i5 processor using a serial implementation in Python using the Gym framework. The relationship between the total time required to generate a specific number of games appears to be linear, as shown in the right image of Figure~\ref{fig:computing-chess}. Actual results are obtained for an average game length of $80$ moves.

\begin{theorem} 
Zermelo's theorem is a game-theory theorem regarding finite two-person games of perfect information in which the players move alternately and the game is not subject to randomization. It indicates that if the game cannot end in a draw, then one of the two players must have a winning strategy~\cite{schwalbe2001zermelo}, i.e. \textit{force a win}. 
\end{theorem}

\begin{remark}
As the chess is a finite two-person games of perfect information in which the players move alternately and the game is not subject to randomization, Zermelo's Theorem can be applied for this game. It states that either White can force a win, or Black can force a win, or both sides can force at least a draw.
\end{remark}

\begin{remark}
Previous results show that the game of Losing Chess is a win for White~\cite{watkins2017losing}. The "losing-winning" move is \textbf{e3} for White.
\end{remark}

Due to finite-bounds of the game and the existence of the 50-move rule, the longest chess game could be up to 4851 moves with a total of 132 different possible per move options. Definitely these are just hypothetical situation analyzing the worst-case scenario, as in real games the possibilities are far smaller.

Hence, the total number of chess games would be at most $132^{4851}$. As a finite game can be simulated in constant time, the above estimation, translated in B-L notations for complexity classes, this means that the perfect algorithm for chess should perform in $O(132^{4851})$. Using the properties of Big-O complexity class, we can state that this algorithm will perform in constant time, with $O(1)$ complexity as this number of total number of chess, regardless how big it is, is still a constant. 

In this situation, the B-L asymptotic notations did not provide useful information and the reason is simple: these notations were developed for asymptotic-scaling problems and algorithms, w/o awareness of discrete values. Even though in most cases these notations were helpful, this is probably not the case in this scenario.

Claude Shannon had studied the implications of a brute force solution for solving chess back in the 1950~\cite{shannon1950xxii}, when he introduced the \textbf{Shannon number}, a conservative lower bound of the game-tree complexity of chess. The purpose was to validate that any perfect chess algorithms based on brute-force are impractical.

The proposed Shannon number was equal with $10^{120}$, taking into consideration a typical game  of 80 moves at a rate of $10^3$ possibilities for each pair of white-black moves. Further work showed that, based on an average branching factor of 35 and an average game length of 80, the lower bound for the chess game-tree is around $10^{123}$, as proofed in~\cite{allis1994searching}.

\subsection{Chess in r-Complexity}
As we previously stated, the perfect algorithm for chess is part of the $O(1)$ complexity class, as its input values are finite-bounded. Thus, the associated r-Complexity class would be $O_{1}(c)$, where $c$ in a finite constant. A human-driven calculus of r-Complexity is not feasible, as there are various run-time aspects that are difficult to be taken into consideration and an exact calculus would imply a even greater effort than solving straightforward the chess problem. Thus, we propose an automatic estimation for this algorithm, that has its B-L Complexity known.

The first step was acquiring data on few game-simulation. Using gym framework, we tracked the time-complexity for various number of episodes with different number of steps per each episodes. Using an average game length of $80$, the chess-solving problem becomes a one-parameter problem that involves the total number of episodes to be generated. This value is lower-bounded by the value of $10^{123}$. A brute-force solution for this algorithm would act almost linearly in terms of number of episodes to be generated.
 
Of course, many optimization can reduce the total time by even orders of magnitude. Regardless of the optimization process, for an input of $10^{123}$, the total estimated time using this algorithm would be $0.855 \cdot 10^{123}$ seconds and thus the r-Complexity would be $O_{1}(0.855 \cdot 10^{123} \cdot HZ)$, where $HZ \approx 2.3 \cdot 10^9$. Such a huge time limit is a result of the greatness of the lower bound for chess. Recall that this is by over $40$ orders of magnitude greater that the total estimated number of atoms in the universe.

\emph{The dream of the perfect algorithm}

All means of computing in $2020$ are at an enormous gap from what it would be needed in order to find the perfect algorithm of chess using brute-force solutions. Based on the estimated r-Complexity (i.e. $O_{1}(0.855 \cdot 10^{123} \cdot HZ)$, where $HZ \approx 2.3 \cdot 10^9$) we present a scenario describing the actual computational cost:
Assume that each atom in the universe is the state-of-the-art computing core of a modern processor, that operates at a frequency of 5GHz. Assume that the perfect algorithm of chess defies Amdahl's Law by assuming a theoretical unbounded speedup in the latency of the execution. Assume we have zero latency between inter-process communication. So far, we built an system consisting of $10^{80}$ computing units operating at $HZ_0 \approx 5 \cdot 10^9$, with zero latency that needs to solve an algorithm with associated complexity of $O_{1}(0.855 \cdot 10^{123} \cdot 2.3 \cdot 10^9)$. That is, assuming perfect distribution and zero overhead, each unit would require approx $10^{42}$ seconds. This is far greater than any estimation of the universe lifespan. Note also that the estimate the total number of fundamental particles in the observable universe is $10^{80}$.
  
\section{Conclusion and future work}

We could further extend our philosophical discussion with many more scenarios, but the point is clear: perfectly solving the game of chess is a far too complicated problem, and yet, we humans, with limited computing power, can naturally play the game of chess charmingly well. The paper prepared additional resources for estimating rComplexity for algorithms with known B-L at the online codebase resource\footnote{https://github.com/raresraf/rafMetrics}.

This work introduced the architecture-dependent r-Complexity metric as an extension of the traditional complexity model, offering enhanced comparison mechanisms for algorithm correlation for bounded input dimensions. Also, the r-Complexity model is susceptible to automatic discovery, a feature that facilitates valuable estimations, when an explicit calculus is impractical, by using methods presented in this work. 

The model can provide better observations and improved understanding when analysing computational algorithm complexity, as outlined for matrix multiplication and chess game in this paper. The r-Complexity model is generic and, in future work new metrics may be analyzed using it, such as average response time for a service communicating over computer networks or L1 cache misses for specific compute-intensive algorithms.

\section*{Acknowledgements}

This work was partially supported by the European Union’s Horizon 2020 research and innovation programme under grant agreement No. 786669 (ReAct) and No. 825377 (UNICORE).

\bibliographystyle{plain}
\bibliography{bibliography}

\end{document}